# Highly anisotropic magnetic phase diagram of the ferromagnetic rare-earth diboride HoB$_2$

Takafumi D. Yamamoto[1, *, †], Hiroyuki Takeya[1], Kensei Terashima[1], Akiko T. Saito[1], and Yoshihiko Takano[1, 2]

[1]*National Institute for Materials Science, Ibaraki 305-0047, Japan*
[2]*Graduate School of Pure and Applied Sciences, University of Tsukuba, Ibaraki 305-8577, Japan*

*Contact Author: td_yamamoto@rs.tus.ac.jp

[†]Present affiliation:
*Department of Materials Science and Technology, Tokyo University of Science, Tokyo 125-8585, Japan.*

**Abstract:**

Rare-earth (RE) compounds have been of enormous interest in condensed matter physics as a platform for the exploration of interesting physical phenomena. Here, we report the successful single crystal growth of HoB$_2$, which exhibits a paramagnetic (PM) – ferromagnetic (FM) phase transition at 15 K and another phase transition at 11 K, and the discovery of a highly anisotropic magnetic phase diagram of this FM diboride. Magnetization measurements suggest that the ferromagnetically ordered moments of Ho$^{3+}$ ions are oriented at a direction tilted by 50 degrees from the *ab*-plane at 2K, and they rotate largely toward the *ab*-plane direction upon application of a magnetic field but hardly toward the *c*-axis direction. Heat capacity and electrical resistivity measurements clearly demonstrate that the low-temperature phase transition at 11 K occurs even under high magnetic fields along the *ab*-plane direction, whereas it disappears immediately by magnetic fields along the *c*-axis direction. Moreover, the field-induced crossover phenomenon between FM and PM phases near 15 K is found to be more promoted when applying a magnetic field along the *ab*-plane direction. The resulting magnetic phase diagram reveals that in-plane magnetic anisotropy is predominant in the present system, which contradicts the previous report of a spin reorientation phenomenon toward the *c*-axis direction between 11 K and 15 K. Taken together, the present findings suggest the presence of strong competition between in-plane and out-of-plane magnetic anisotropies in HoB$_2$, giving rise to the unique FM spin arrangement.

# I. INTRODUCTION

Rare-earth (RE) borides, a wide class of inorganic compounds, have offered an attractive playground for the exploration of intriguing electronic and magnetic properties [1-3]. For example, cubic $REB_6$ (RE = La–Lu) compounds, possibly the most studied system thus far, is a showcase for dense Kondo behavior [4, 5], Kondo insulators [6, 7], quadrupolar ordering phenomena [8–10], and low-carrier density ferromagnetism with colossal magnetoresistance [11]. In tetragonal $REB_4$ (RE = La–Lu) compounds, which has recently attracted attention as a frustrated Shastry-Sutherland system, one can encounter complex magnetic phase diagrams with fractionalized magnetization plateau regimes [12–14] and non-collinear or incommensurate magnetic phases [14, 15]. The diverse physical properties of these RE borides originate from complex competition between various interactions, including not only major interactions such as the so-called RKKY magnetic interactions, Kondo effects, and crystalline electric field (CEF) effects, but also weaker interactions such as magnetoelastic couplings and quadrupolar interactions.

Layered $REB_2$ (RE = Gd–Lu) compounds with an $AlB_2$-type hexagonal structure of space group $P6/mmm$ (see Fig. S1 in the Supplemental Material [16]) have not been studied as much as other RE borides despite the unique feature among RE borides that most of the series of compounds exhibit ferromagnetism [1, 3]. This ferromagnetism has been widely attributed to a conventional paramagnetic (PM) – ferromagnetic (FM) phase transition with a Curie temperature $T_C$ of 7.2–151 K [1, 17, 18], while a previous neutron diffraction experiment pointed out a possible complex magnetic structure than a collinear FM order [19]. More interestingly, some of $REB_2$ compounds (RE = Tb and Dy) have been reported to undergo another phase transition at lower temperatures following the PM–FM transition [1, 20], but the origin of which has not been clarified yet.

Recently, it was discovered that $HoB_2$ also exhibits not only a second order PM–FM transition at $T_C$ of 15 K but also a low-temperature phase transition at $T_{low}$ = 11 K [21]. One of the significant findings regarding this phase transition is a spin reorientation phenomenon revealed by previous neutron powder diffraction (NPD) experiments on $HoB_2$ [22]: as the temperature drops from $T_C$ to $T_{low}$, the ferromagnetically ordered moments of $Ho^{3+}$ ions gradually change their orientation from the *ab*-plane direction toward the *c*-axis direction, followed by tilting 50° from the *ab*-plane below $T_{low}$. Another interesting finding is the robustness of specific heat peaks to a magnetic field [21, 22]: in contrast to the rapid disappearance of the peak near $T_C$ upon the application of a magnetic field, the peak near $T_{low}$ is clearly visible even at 5 T. These observations imply that the low-temperature phase transition in $HoB_2$ is not a simple magnetic phase transition in which the magnetic interaction is solely involved.

To gain more insight into the low-temperature phase transition of $REB_2$, it would be of importance to investigate the physical properties of single crystals, especially to clarify the magnetic anisotropy. However, to the best of our knowledge, there has been no report thus far on the single crystal growth in $REB_2$ (RE = Tb, Dy, and Ho). Although Zhou et al. have successfully synthesized preferred-oriented samples by an arc-melting method [23], the X-ray diffraction (XRD) patterns for a certain direction of their samples show both (00*l*) and (*h*00) reflections, indicating the imperfect crystal orientation of these samples.

In this study, we succeeded in synthesizing single crystals of HoB$_2$ and revealed the magnetic anisotropy of this diboride through thermodynamic and electrical resistivity measurements. We find that the easy axis of magnetization is indeed tilted by 50° from the *ab*-plane at 2 K, as proposed by the previous NPD experiment [22]. Interestingly, the ferromagnetically ordered moments rotate largely toward the *ab*-plane direction upon the application of a magnetic field, but hardly toward the *c*-axis direction. The most remarkable finding is the anisotropic magnetic field effect on the low-temperature phase transition: it occurs even under high magnetic fields along the *ab*-plane direction, whereas disappears immediately with magnetic fields along the *c*-axis direction. Furthermore, a field-induced crossover phenomenon between FM and PM phases near $T_C$ is found to be more promoted when a magnetic field is applied along the *ab*-plane direction, though it can be observed regardless of the field direction. A highly anisotropic magnetic phase diagram is established for HoB$_2$, revealing that in-plane magnetic anisotropy is predominant in the present system. The presence of competing magnetic anisotropies in HoB$_2$ is discussed.

## II. EXPERIMENTAL DETAILS

Black, plate-like single crystals of HoB$_2$, shown in Fig. 1(a), were synthesized from a large amount (~100 g) of arc-melts of a stoichiometric mixture of Ho (3N) and B (3N) by a conventional arc-melting method under an argon atmosphere. Other preparation methods, such as a usual floating-zone method and a flux method using aluminum flux, were also attempted, but only HoB$_4$ single crystals were obtained. The crystallinity and crystallographic orientation of HoB$_2$ single crystal were evaluated by back-reflection X-ray Laue photographs with a tungsten target. As shown in Fig. 1(b), the Laue diffraction pattern along the <001> axis represents a six-fold symmetric pattern with clear circular Laue spots. Moreover, room-temperature powder XRD measurements using a Rigaku X-ray diffractometer (Cu *K*α radiation) indicate that the powdered single-crystal sample is in a single phase [see Fig. 1(c)], except for a tiny amount of Ho$_2$O$_3$ oxide that may be introduced during the powdering process for the measurement. These results confirm the high quality of single crystals obtained in this study.

For physical property measurements, a HoB$_2$ single crystal was formed into a rectangular shape with dimensions of ~2.0 mm × ~0.6 mm × ~0.2 mm, where the shortest side is along the crystallographic *c*-axis. Magnetization (*M*) data were acquired using a Quantum Design SQUID magnetometer for external magnetic fields ($H_{ex}$) along the *ab*-plane direction, the *c*-axis direction, and the direction tilted by 50° from the *ab*-plane. Hereafter, each case is referred to as $\mu_0H_{ex}$ // *ab*, $\mu_0H_{ex}$ // *c*, and $\mu_0H_{ex}$ // 50°, respectively. Specific heat (*C*) and electrical resistivity (ρ) were measured for $\mu_0H_{ex}$ // *ab* and $\mu_0H_{ex}$ // *c* using a thermal relaxation method and a four-terminal method in a Quantum Design PPMS. The resistivity measurements were performed in the so-called longitudinal configuration for $\mu_0H_{ex}$ // *ab* and in the transverse configuration for $\mu_0H_{ex}$ // *c*, where the electrical current (*I*) of 2.5 mA was applied along the *ab*-plane direction.

Although the sample geometry was imperfectly rectangular, the demagnetization effects on the sample used for the measurements were evaluated for $\mu_0 H_{ex}$ // $ab$ and $\mu_0 H_{ex}$ // $c$ by estimating demagnetization factors based on an approximate formula for rectangular samples [24]. Figure S2 shows the ratio of internal magnetic field to external magnetic field ($H_{int}$ / $H_{ex}$) as a function of temperature [16], indicating that the demagnetization effect on our sample is strongly temperature-dependent between 10 K and 20 K; the lower the $\mu_0 H_{ex}$, the more pronounced this trend becomes. Furthermore, the demagnetization effect is noticeable especially for $\mu_0 H_{ex}$ // $c$: for instance, $H_{int}$ is decreased to about 10% of $\mu_0 H_{ex}$ = 1 T at 10 K [see Fig. S2(b)]. In the following, $H_{ex}$ is used to denote the magnetic field for clarity, and the influence of demagnetization corrections on the results will be explained accordingly.

### III. RESULTS AND DISCUSSION

We first examine the magnetic anisotropy of HoB$_2$ single crystal at the lowest temperature. Figure 2 shows the magnetization $M$ at 2 K plotted as a function of external magnetic fields for $\mu_0 H_{ex}$ // $ab$, $\mu_0 H_{ex}$ // $c$, and $\mu_0 H_{ex}$ // 50°, respectively. In all three cases, $M$ rises at low magnetic fields below 1 T and tends to saturate at high fields above 3 T, where the magnitude is largest for $\mu_0 H_{ex}$ // 50°, followed in order by for $\mu_0 H_{ex}$ // $ab$ and for $\mu_0 H_{ex}$ // $c$. Note that the demagnetization correction only shifts each data laterally toward $\mu_0 H_{ex}$ = 0 T, so the magnitude relation of $M$ at high magnetic fields is unlikely to be changed. Given that the magnetic structure proposed by previous NPD experiments is correct, *i.e.*, the ordered magnetic moments are ferromagnetically aligned in a direction tilted 50° from the *ab*-plane at 2 K [22], $M$ is expected to be maximal when a magnetic field is applied in this direction. Our magnetization data are consistent with this expectation, suggesting that the easy axis of magnetization is indeed tilted by 50° from the *ab*-plane. The exact verification of which requires further experiments using a sample rotator.

For a more quantitative discussion, we compare the values of magnetization at 5 T ($M_{5T}$) for three magnetic field directions: 9.01 $\mu_B$/Ho for $\mu_0 H_{ex}$ // 50°, 7.92 $\mu_B$/Ho for $\mu_0 H_{ex}$ // $ab$, and 6.86 $\mu_B$/Ho for $\mu_0 H_{ex}$ // $c$, respectively. From the above considerations, one can assume that the ferromagnetically ordered moments with a magnitude of 9.01 $\mu_B$/Ho are oriented at a direction tilted by 50° from the *ab*-plane. This magnitude is smaller than the expected value for the free Ho$^{3+}$ ions (10 $\mu_B$/Ho), implying the significance of the CEF effect in HoB$_2$. Notably, as can be seen from the inset of Fig. 2, the directional cosine of 9.01 $\mu_B$/Ho with respect to the *c*-axis, 6.90 $\mu_B$/Ho, is in good agreement with the $M_{5T}$ value for $\mu_0 H_{ex}$ // $c$, suggesting that the magnetic moments hardly rotate when a magnetic field is applied along the *c*-axis direction [25]. On the contrary, the $M_{5T}$ value for $\mu_0 H_{ex}$ // $ab$ is much larger than the directional cosine with respect to the *ab*-plane (5.79 $\mu_B$/Ho), representing the relatively large rotation of magnetic moments upon the application of a magnetic field along the *ab*-plane direction.

Unlike the magnetization at 2 K, the magnetic susceptibility $M/H_{ex}$ at higher temperatures has similar features regardless of the field direction (see Fig. S3 in the Supplemental Material [16]). With increasing temperature, $M/H_{ex}$ at 0.01 T exhibits a cusp anomaly around $T_{low}$ = 11 K and then a gradual decrease, followed by a sudden drop above $T_C$ = 15 K. In addition, as shown in Fig. S4, the overall features of iso-field magnetization (*M-T*) curves at high magnetic fields are also similar between 2 K and 50 K for all the cases [16]. As such, the magnetic anisotropy of HoB$_2$ is not evident in the magnetic properties at the high temperature side.

On the other hand, the temperature dependences of $C$ and $\rho$ for $\mu_0 H_{ex}$ // *ab* and $\mu_0 H_{ex}$ // *c*, shown in Figs. 3(a)–3(d), clearly demonstrate anisotropic field effects on the two phase transitions in HoB$_2$. At zero magnetic field, a sharp peak in $C$ and a clear kink in $\rho$ are observed around $T_{low}$, being ascribed to the occurrence of the low-temperature phase transition. Interestingly, these anomalies are obvious under high magnetic fields for $\mu_0 H_{ex}$ // *ab* [Figs. 3(a) and 3(b)], while they become obscured for $\mu_0 H_{ex}$ // *c* [Figs. 3(c) and 3(d)]. Moreover, the peak and kink temperatures shift to the higher (lower) temperature side with increasing magnetic field for $\mu_0 H_{ex}$ // *ab* ($\mu_0 H_{ex}$ // *c*), which is more visible in Figs. S5 [16]. These results indicate that the low-temperature phase transition occurs even under high magnetic fields along the *ab*-plane direction, whereas it disappears immediately by magnetic fields along the *c*-axis direction.

The magnetic field effect on the PM–FM phase transition is somewhat different. When a magnetic field is applied, the broadening of specific heat peak and resistivity kink near $T_C$ is observed for both cases, representing a crossover phenomenon between PM and FM phases. Since the magnetic field favors FM orders regardless of their direction, it is reasonable that this phenomenon is observe for either field direction. However, these anomalies are broadened and shift to higher temperatures more significantly for $\mu_0 H_{ex}$ // *ab* than for $\mu_0 H_{ex}$ // *c*, suggesting that magnetic anisotropy is substantial even at the temperature near $T_C$. The differences due to the magnetic field directions are still observed even if the internal magnetic field is the same, *e.g.*, when comparing the 1 T data for $\mu_0 H_{ex}$ // *c* with the 0.6 T data for $\mu_0 H_{ex}$ // *ab* (see Figs. S2 and S6 [16]).

Figures 4(a) and 4(b) show the $\mu_0 H_{ex} - T$ phase diagrams of HoB$_2$ for $\mu_0 H_{ex}$ // *ab* and $\mu_0 H_{ex}$ // *c*, established based on the temperature and magnetic field dependence of $C$ and $\rho$. The details of the determination of the phase diagram are described in the Supplemental Material (see also Figs. S7 and S8 [16]). At zero magnetic field, three distinct phases can be found: a PM phase above $T_C$ (phase I), a FM phase between $T_{low}$ and $T_C$ (phase II), and another FM phase below $T_{low}$ (phase III). It is generally understood that a conventional PM–FM phase transition is removed by applying a magnetic field, owing to the energetic advantage of the magnetic moment being aligned along the field direction [26]. Hence, no phase boundary for PM–FM transition is present at any finite field. Instead, the crossover lines separating phase I and phase II are represented near $T_C$ in Figs. 4(a) and 4(b), respectively. Both two lines do not coincide with each other, reflecting the magnetic anisotropy as discussed above. Remarkably, the magnetic phase diagram is highly anisotropic with respect to the low-temperature phase transition. For $\mu_0 H_{ex}$ // *ab* [Fig. 4(a)], phase III exists over a wide field range from 0 to 5 T and even extends slightly into the higher temperature region at high magnetic fields. Conversely, as manifested in the decrease in $T_{low}$, the phase III region shrinks with applying a magnetic field for $\mu_0 H_{ex}$ // *c* [Fig. 4(b)]. Furthermore, phase III seems to completely vanish at $\mu_0 H_{ex} = \sim 1.7$ T at 0 K, which is deduced from the dotted line in Fig. 4(b). It is should be noted here that since the demagnetization effect for $\mu_0 H_{ex}$ // $c$ is pronounced at low temperatures, the phase III may actually be collapsed by lower magnetic fields below 0.5 T.

      The magnetic phase diagrams in Fig. 4 suggest that phase III becomes energetically stable (unstable) when the magnetic moments are aligned along the $ab$-plane ($c$-axis) direction by applying a magnetic field. From this point of view, in-plane magnetic anisotropy is considered to be favorable for phase III, which is consistent with the large rotation of magnetic moments at 2 K for $\mu_0 H_{ex}$ // $ab$ as discussed above. Moreover, the crossover line near $T_C$ extends to a higher temperature side for $\mu_0 H_{ex}$ // $ab$ than $\mu_0 H_{ex}$ // $c$, representing that the development of the FM order (*i.e.*, phase II) is more promoted by magnetic fields along the $ab$-plane direction rather than the $c$-axis direction. These facts indicate that in-plane magnetic anisotropy is likely to be predominant in both phase II and phase III. However, as observed in the previous NPD experiment at zero magnetic field [22], HoB$_2$ experiences the spin reorientation phenomenon toward the $c$-axis direction between $T_C$ and $T_{low}$, which clearly contradicts with the present findings. Such a discrepancy suggests the presence of competition for magnetic anisotropy in this system. Indeed, considering the fact that the angle between the $ab$-plane and the ordered magnetic moment at 5 T and 2 K for $\mu_0 H_{ex}$ // $ab$ is estimated to be 28° from the ratio of $M_{5T}$ value for $\mu_0 H_{ex}$ // $ab$ to that for $\mu_0 H_{ex}$ // 50° (Fig. 2), it seems that out-of-plane magnetic anisotropy prevents the complete rotation of magnetic moments along the $ab$-plane direction. On the other hand, no rotation of magnetic moments toward the $c$-axis direction by high magnetic fields at 2 K for $\mu_0 H_{ex}$ // $c$ implies the influence of in-plane magnetic anisotropy even in this region where the phase III is no longer present. To sum up the above, we propose that there is strong competition between in-plane and out-of-plane magnetic anisotropies in HoB$_2$, giving rise to the unique spin arrangement tilted by 50° from the $ab$-plane below 11 K.

      The origin of the low-temperature phase transition in HoB$_2$ is still unknown at this stage, but several previous studies may provide clues to consider its driving force. Terada et al. [27] have proposed the CEF level scheme in HoB$_2$ from the fitting of inelastic NPD data, revealing that the CEF ground state is represented as $|J_z = 0\rangle$ by using the eigenstate of $J_z$, where the $J_z$ is the $z$-component of total angular momentum $J$. Notably, the first excited states with a small energy gap of $\sim 0.93$ meV include two nearly degenerated doublet states, $|J_z = \pm 1\rangle$ and $|J_z = \pm 7\rangle$, implying a possible competition between in-plane and out-of-plane magnetic anisotropies at low temperatures. Meanwhile, other minor factors besides magnetic interactions and CEF effects may also be important. Gschneidner et al. [28–32] have pointed out the importance of competing magnetoelastic and quadrupolar interactions in the emergence of multiple magnetic ordering phenomena in REAl$_2$ alloys. Near the low-temperature phase transition of these compounds, the *M-T* curve under low magnetic fields exhibits an unusual thermal hysteresis between field-cooling-cooling (FCC) and field-cooling-warming (FCW) process within a FM phase (e.g., see Fig. 2 in Ref. 32): The FCC magnetization is higher than the FCW magnetization. It is interesting to note that as shown in Fig. S9 [16], a similar magnetization curve is observed in HoB$_2$

single crystals. In order to fully elucidate the magnetism in HoB$_2$, it is desirable to determine CEF levels of Ho$^{3+}$ ions by performing CEF calculations based on the low-temperature crystal structure and comparing the results with thermodynamic measurement data of single crystals.

**IV. SUMMARY**

In summary, we investigated the magnetic anisotropy of the rare-earth diboride HoB$_2$, which exhibits a paramagnetic (PM) – ferromagnetic (FM) phase transition at 15 K and another phase transition at 11 K, through thermodynamic and transport measurements on single crystals. Magnetization at 2 K is largest when a magnetic field is applied along a direction tilted by 50 degrees from the *ab*-plane, suggesting that the easy axis of magnetization is oriented at this direction at the lowest temperature, as proposed by the previous powder neutron diffraction. Interestingly, the ferromagnetically ordered moments of Ho$^{3+}$ ions are found to rotate largely toward the *ab*-plane direction upon the application of magnetic field, but hardly toward the *c*-axis direction. Heat capacity and electrical resistivity measurements clearly demonstrate that the low-temperature phase transition at 11 K occurs even under high magnetic fields along the *ab*-plane direction whereas it disappears immediately by magnetic fields along the *c*-axis direction. Furthermore, the field-induced crossover phenomenon between FM and PM phases near 15 K is more promoted by applying a magnetic field along the *ab*-plane direction, though it is observed regardless of the field direction. The above results unveil a highly anisotropic magnetic phase diagram of HoB$_2$, revealing that in-plane magnetic anisotropy is predominant in the present system, which contradicts with the previous report of spin reorientation phenomenon toward the *c*-axis direction between 11 K and 15 K. Taken together, the present findings suggest that there is strong competition between in-plane and out-of-plane magnetic anisotropies in HoB$_2$, giving rise to the unique spin arrangement tilted by 50 degrees from the *ab*-plane below 11 K. Finally, this study is the first report of a magnetic phase diagram in rare-earth diborides (REB$_2$), exemplifying the importance of investigations on single crystals in clarifying unique physical properties of REB$_2$. Similar studies in other REB$_2$ may lead to further discovery of novel physical phenomena in rare-earth compounds.

**ACKNOWLEDGEMENTS**

This work was supported by JST-Mirai Program Grant Number JPMJMI18A3, Japan.

**REFERENCES**

[1] K. H. J. Buschow, *Boron and Refractory Borides* (Ed: V. I. Matkovich), Springer, Berlin, Heidelberg, pp. 494-515 (1977).

[2] J. Etourneau and P. Hagenmuller, Structure and physical features of the rare-earth borides, Philos. Mag. B **52**, 589 (1985).

[3] S. Gabani, K. Flachbart, K. Siemensmeyer, and T. Mori, Magnetism and superconductivity of rare earth borides, J. Alloys Compd. **82**, 153201 (2020).

[4] N. Sato, A. Sumiyama, S. Kunii, H. Nagano, and T. Kasuya, Interaction between Kondo states and the Hall Effect of dense Kondo system $Ce_xLa_{1-x}B_6$, J. Phys. Soc. Jpn. **54**, 1923 (1985).

[5] S. Nakamura, T. Goto, and S. Kunii, Magnetic phase diagrams of the dense Kondo compounds $CeB_6$ and $Ce_{0.5}La_{0.5}B_6$, J. Phys. Soc. Jpn. **64**, 3941 (1995).

[6] J. W. Allen, R. M. Martin, B. Batlogg, P. Wachter, Mixed valent $SmB_6$ and gold-SmS: Metals or insulators?, J. Appl. Phys. **49**, 2078 (1978).

[7] J. Ying, L. Tang, F. Chen, X. Chen, and V. V. Struzhkin, Coexistence of metallic and insulating channels in compressed $YbB_6$, Phys. Rev. B **97**, 121101 (R) (2018).

[8] J. M. Effantin, J. Rossat-Mignod, P. Burlet, H. Bartholin, S. Kunii, and T. Kasuya, Magnetic phase diagram of $CeB_6$, J. Magn. Magn. Mater. **47–48**, 145 (1985).

[9] S. Kobayashi, M. Sera, M. Hiroi, T. Nishizaki, N. Kobayashi, and S. Kunii, Anisotropic magnetic phase diagram of $PrB_6$ dominated by the $O_{xy}$ antiferro-quadrupolar interaction, J. Phys. Soc. Jpn. **70**, 1721 (2001).

[10] T. Goto, Y. Nemoto, Y. Nakano, S. Nakamura, T. Kajitani, and S. Kunii, Quadrupolar effect of $HoB_6$ and $DyB_6$, Physica B **281–282**, 586 (2000).

[11] S. Süllow, I. Prasad, M. C. Aronson, S. Bogdanovich, J. L. Sarrao, and Z. Fisk, Metallization and magnetic order in $EuB_6$, Phys. Rev. B 62, 11626 (2000).

[12] K. Siemensmeyer, E. Wulf, H.-J. Mikeska, K. Flachbart, S. Gabáni, S. Mat'aš, P. Priputen, A. Efdokimova, and N. Shitsevalova, Fractional magnetization plateaus and magnetic order in the Shastry-Sutherland magnet $TmB_4$, Phys. Rev. Lett. **101**, 177201 (2008).

[13] S. Yoshii, T. Yamamoto, M. Hagiwara, S. Michimura, A. Shigekawa, F. Iga, T. Takabatake, and K. Kindo, Multistep magnetization plateaus in the Shastry-Sutherland system $TbB_4$, Phys. Rev. Lett. **101**, 087202 (2008).

[14] D. Brunt, G. Balakrishnan, D. A. Mayoh, M. R. Lees, D. Gorbunov, N. Qureshi, and O. A. Petrenko, Magnetisation process in the rare earth tetraborides, $NdB_4$ and $HoB_4$, Sci. Rep. **8**, 232 (2018).

[15] D. Brunt, G. Balakrishnan, A. R. Wildes, B. Ouladdiaf, N. Qureshi, and O. A. Petrenko, Field-induced magnetic states in holmium tetraboride, Phys. Rev. B **95**, 024410 (2017).

[16] See the Supplemental Material at [URL] for crystal structure, demagnetization effect, temperature dependence of the magnetic properties at high temperatures, detailed magnetic field dependence of

specific heat and electrical resistivity, determination of the magnetic phase diagram, and thermal hysteresis of iso-field magnetization curves.

[17] T. Mori, T. Takimoto, A. L.- Jasper, R. C.- Gil, W. Schnelle, G. Auffermann, H. Rosner, and Y. Grin, Ferromagnetism and electronic structure of $TmB_2$, Phys. Rev. B **79**, 104418 (2009).

[18] M. A. Avila, S. L. Bud'ko, C. Petrovic, R. A. Ribeiro, P. C. Canfield, A. V. Tsvyashchenko, and L. N. Fomicheva, Synthesis and properties of $YbB_2$, J. Alloys Compd. **358**, 56 (2003).

[19] G. Will and W. Schäfer, Neutron diffraction and the magnetic structures of some rare earth diborides and tetraborides, J. Less-Common Met. **67**, 31 (1979).

[20] H. Meng, B. Li, Z. Han, Y. Zhang, X. Wang, and Z. Zhang, Reversible magnetocaloric effect and refrigeration capacity enhanced by two successive magnetic transitions in $DyB_2$, Sci. China Technol. Sci. **55**, 501 (2012).

[21] P. B. Castro, K. Terashima, T. D. Yamamoto, Z. Hou, S. Iwasaki, R. Matsumoto, S. Adachi, Y. Saito, P. Song, H. Takeya, and Y. Takano, Machine-learning-guided discovery of the gigantic magnetocaloric effect in $HoB_2$ near the hydrogen liquefaction temperature, NPG Asia Mater. **12**, 35 (2020).

[22] N. Terada, K. Terashima, P. B. Castro, C. V. Colin, H. Mamiya, T. D. Yamamoto, H. Takeya, O. Sakai, Y. Takano, and H. Kitazawa, Relationship between magnetic ordering and gigantic magnetocaloric effect in $HoB_2$ studied by neutron diffraction experiment, Phys. Rev. B **102**, 094435 (2020).

[23] X. Zhou, Y. Shan, T. Luo, Y. Peng, and H. Fu, Large rotating magnetocaloric effect of textured polycrystalline $HoB_2$ alloy contributed by anisotropic ferromagnetic susceptibility, Appl. Phys. Lett. **120**, 132401 (2022).

[24] A. Aharoni, Demagnetizing factors for rectangular ferromagnetic prisms, J. Appl. Phys. **83**, 3432 (1998).

[25] Y. Higuchi, H. Sugawara, Y. Aoki, and H. Sato, Anisotropic magnetization in $DyCo_2$ single crystal, J. Phys. Soc. Jpn. **69**, 4114 (2000).

[26] S. Blundell, *Magnetism in Condensed Matter* (Oxford University Press Inc., New York, 2001).

[27] N. Terada, H. Mamiya, H. Saito, T. Nakajima, T. D. Yamamoto, K. Terashima, H. Takeya, O. Sakai, S. Itoh, Y. Takano, M. Hase, and H. Kitazawa, Crystal electric field level scheme leading to giant magnetocaloric effect for hydrogen liquefaction, Commun. Mater. **4**, 13 (2023).

[28] A. L. Lima, K. A. Gschneidner, Jr., V. K. Pecharsky, and A. O. Pecharsky, Disappearance and reappearance of magnetic ordering upon lanthanide substitution in $(Er_{1-x}Dy_x)Al_2$, Phys. Rev. B **68**, 134409 (2003).

[29] R. Nirmala, Y. Mudryk, V. K. Pecharsky, and K. A. Gschneidner, Jr., Unusual magnetism of $Er_{0.75}Dy_{0.25}Al_2$, Phys. Rev. B **76**, 014407 (2007).

[30] M. Khan, K. A. Gschneidner, Jr., and V. K. Pecharsky, Multiple magnetic ordering phenomena evaluated by heat capacity measurements in $Er_{1-x}Tb_xAl_2$ Laves-phase alloys, Phys. Rev. B **80**, 224408 (2009).

[31] R. Nirmala, D. Paudyal, V. K. Pecharsky, and K. A. Gschneidner, Jr., Magnetic properties of Er$_{1-x}$Dy$_x$Al$_2$ ($0 \leq x \leq 1$) compounds in low applied fields, J. Appl. Phys. **107**, 09A723 (2010).

[32] M. Khan, Y. Mudryk, D. Paudyal, K. A. Gschneidner, Jr., and V. K. Pecharsky, Experimental and theoretical study of the magnetic and structural properties of Er$_{0.75}$Tb$_{0.25}$Al$_2$, Phys. Rev. B **82**, 064421 (2010).

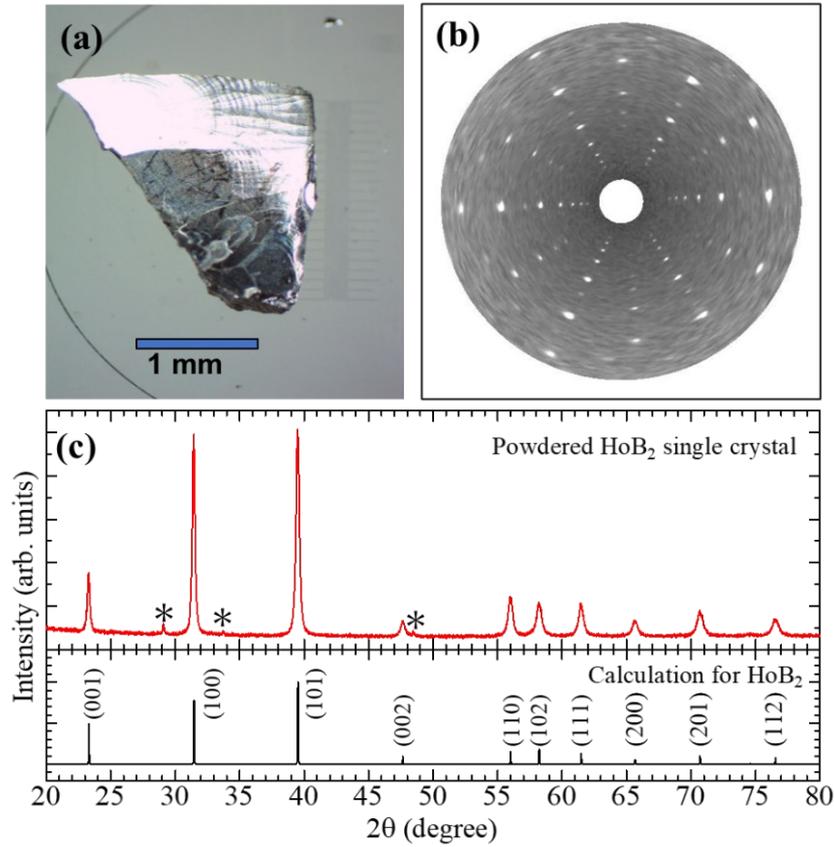

Fig. 1. (a) Photograph of a synthesized single crystal of HoB$_2$. (b) Back-reflection X-ray Laue diffraction pattern along the sixfold-symmetric <001> axis. (c) Room-temperature powder XRD pattern of the powdered HoB$_2$ single crystal (the upper panel) and the calculated pattern for HoB$_2$ (the lower panel). Diffraction peaks from impurity Ho$_2$O$_3$ are marked with an asterisk (*).

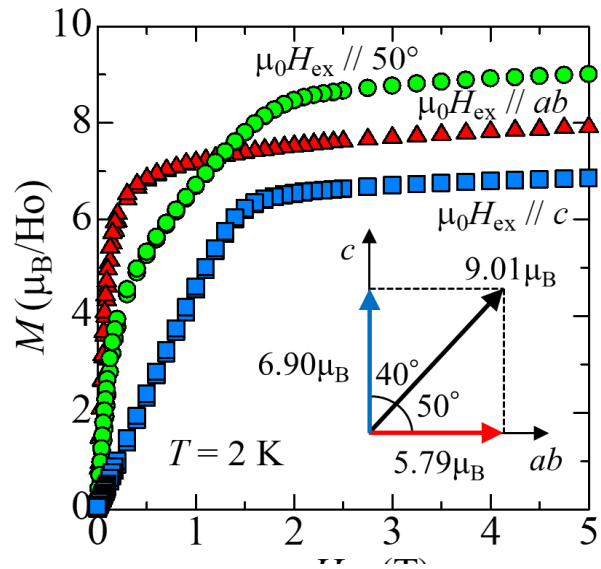

Fig. 2. Magnetization $M$ of the HoB$_2$ single crystal at 2 K plotted as a function of an external magnetic field $\mu_0 H_{ex}$ for $\mu_0 H_{ex}$ // $ab$, $\mu_0 H_{ex}$ // $c$, and $\mu_0 H_{ex}$ // 50°, respectively (see the text). The inset shows the projection of the $M$ value at 5 T for $\mu_0 H_{ex}$ // 50° to the $ab$-plane and $c$-axis directions, respectively.

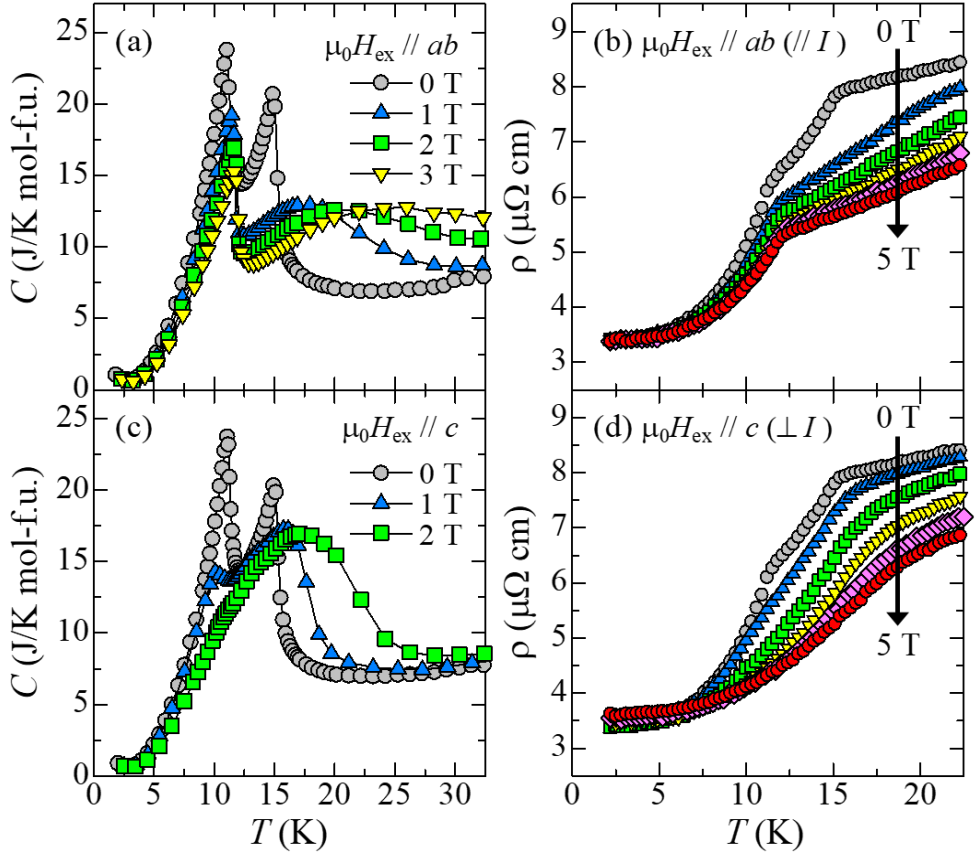

Fig. 3. Temperature dependence of the specific heat $C$ and the electrical resistivity $\rho$ under several external magnetic fields up to 5 T for (a), (b) $\mu_0 H_{ex}$ // $ab$ and (c), (d) $\mu_0 H_{ex}$ // $c$, respectively. The $\mu_0 H_{ex}$ increment in (b) and (d) is 1 T-step for each.

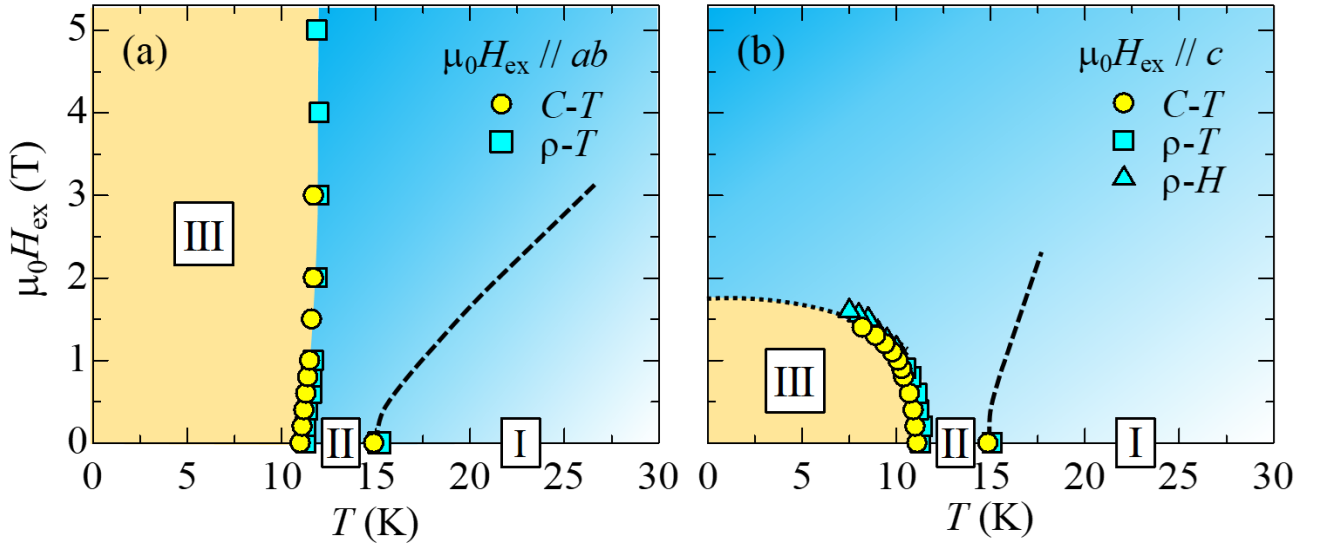

Fig. 4. The $\mu_0 H_{ex} - T$ phase diagrams of HoB$_2$ for (a) $\mu_0 H_{ex}$ // $ab$ and (b) $\mu_0 H_{ex}$ // $c$, where phase I is the paramagnetic (PM) phase; phase II is the ferromagnetic (FM) phase between 11 K and 15 K; phase III is the FM phase below 11 K. The transition temperatures and critical fields are determined from the temperature dependence of the specific heat ($C$-$T$) and resistivity ($\rho$-$T$), and the field dependence of resistivity ($\rho$-$H$). The dashed lines depict the crossover line separating FM and PM phases (see the text). The dotted line in (b) is a guide to the eye.

# Supplemental Material for highly anisotropic magnetic phase diagram of the ferromagnetic rare-earth diboride HoB$_2$

Takafumi D. Yamamoto[1,†], Hiroyuki Takeya[1], Kensei Terashima[1], Akiko T. Saito[1], Takenori Numazawa[1], and Yoshihiko Takano[1,2]

[1]*National Institute for Materials Science, 1-2-1 Sengen, Tsukuba, Ibaraki 305-0047, Japan*
[2]*Graduate School of Pure and Applied Sciences, University of Tsukuba, 1-1-1 Tennodai, Tsukuba, Ibaraki 305-8577, Japan*

[†]Present affiliation:
*Department of Materials Science and Technology, Tokyo University of Science, Tokyo 125-8585, Japan.*

**CONTENTS**

# I. CRYSTAL STRUCTURE OF REB$_2$ COMPOUNDS

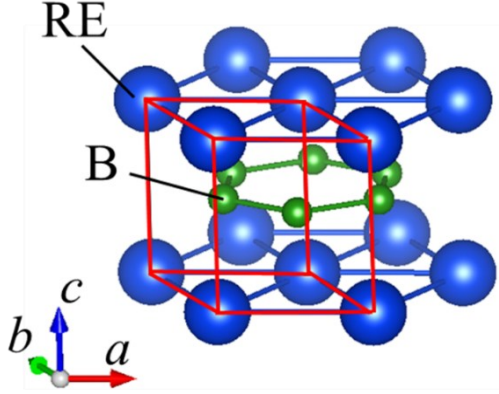

Fig. S1. Crystal structure of REB$_2$ compounds, illustrated by using the VESTA 3 software [K. Momma and F. Izumi, J. Appl. Crystallogr. **44**, 1272 (2011).]. The RE-layers forming the triangular lattice and the B-layers forming the honeycomb lattice are stacked alternately along the *c*-axis. The unit cell is represented by the red solid lines.

# II. EVALUATION OF DEMAGNETIZATION EFFECTS

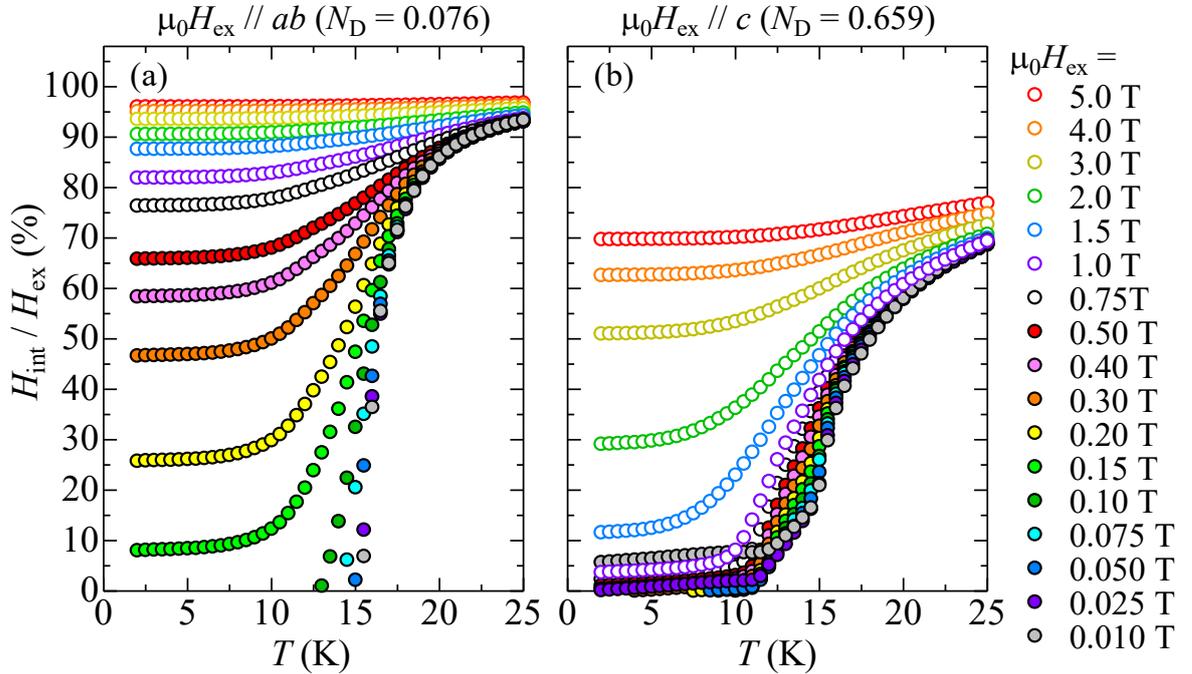

Fig. S2. The ratio of internal magnetic field $H_{int}$ to each external magnetic field $H_{ex}$ as a function of temperature for (a) $\mu_0 H_{ex}$ // *ab* and (b) $\mu_0 H_{ex}$ // *c*. $H_{int}$ was calculated by using the formula $H_{int} = H_{ex} - N_D \rho M(T, H_{ex})$, where $N_D$ the demagnetization factor, $\rho$ the density, and $M(T, H_{ex})$ the measured mass magnetization. The iso-field magnetization curves plotted in Fig. S4 were used for the calculation of $H_{int} / H_{ex}$.

## III. MAGNETIC PROPERTIES OF HoB$_2$ AT HIGH TEMPERATURES

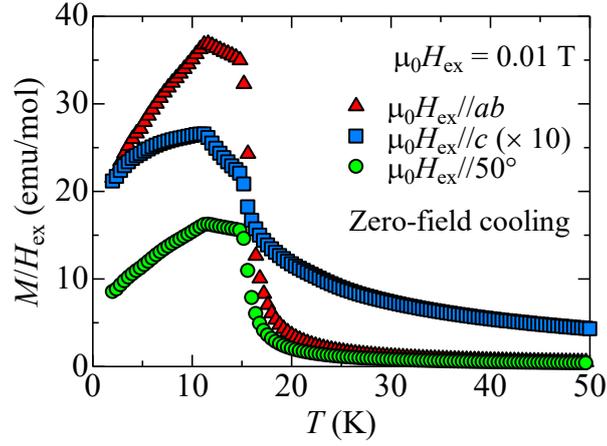

Fig. S3. Temperature dependence of the magnetic susceptibility $M / H_{ex}$ at $\mu_0 H_{ex} = 0.01$ T for $\mu_0 H_{ex} \, // \, ab$, $\mu_0 H_{ex} \, // \, c$, and $\mu_0 H_{ex} \, // \, 50°$ (see the main text), measured in warming after zero-field cooling processes. The data for $\mu_0 H_{ex} \, // \, c$ is shown multiplied by a factor of 10 for easy comparison.

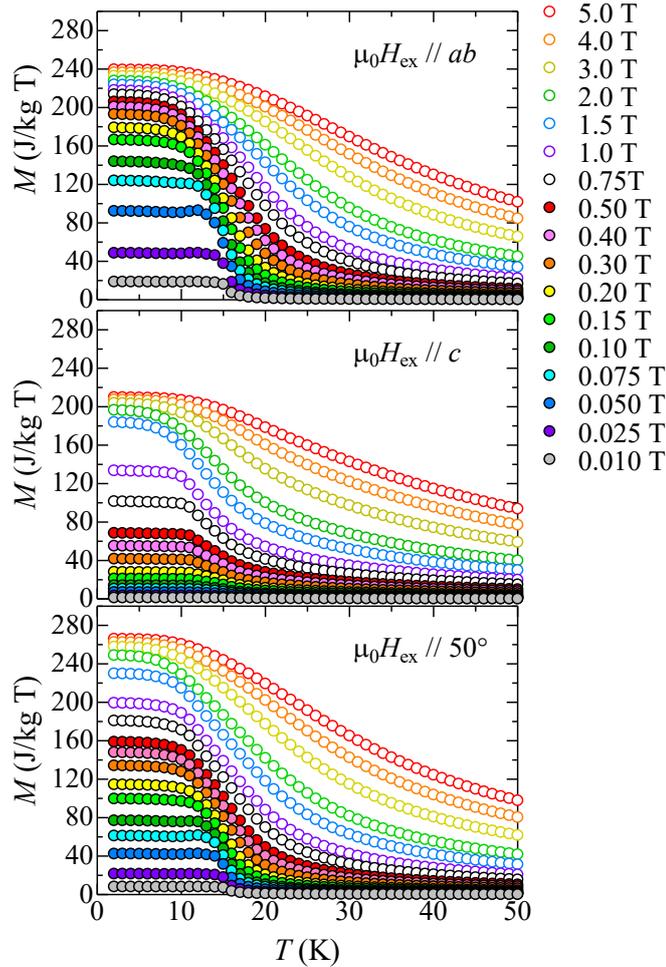

Fig. S4. Iso-field magnetization curves under various external magnetic fields $\mu_0 H_{ex}$ for $\mu_0 H_{ex} \, // \, ab$, $\mu_0 H_{ex} \, // \, c$, and $\mu_0 H_{ex} \, // \, 50°$.

# IV. MAGNETIC FIELD DEPENDENCE OF SPECIFIC HEAT AND ELECTRICAL RESISTIVITY NEAR THE LOW-TEMPERATURE PHASE TRANSITION

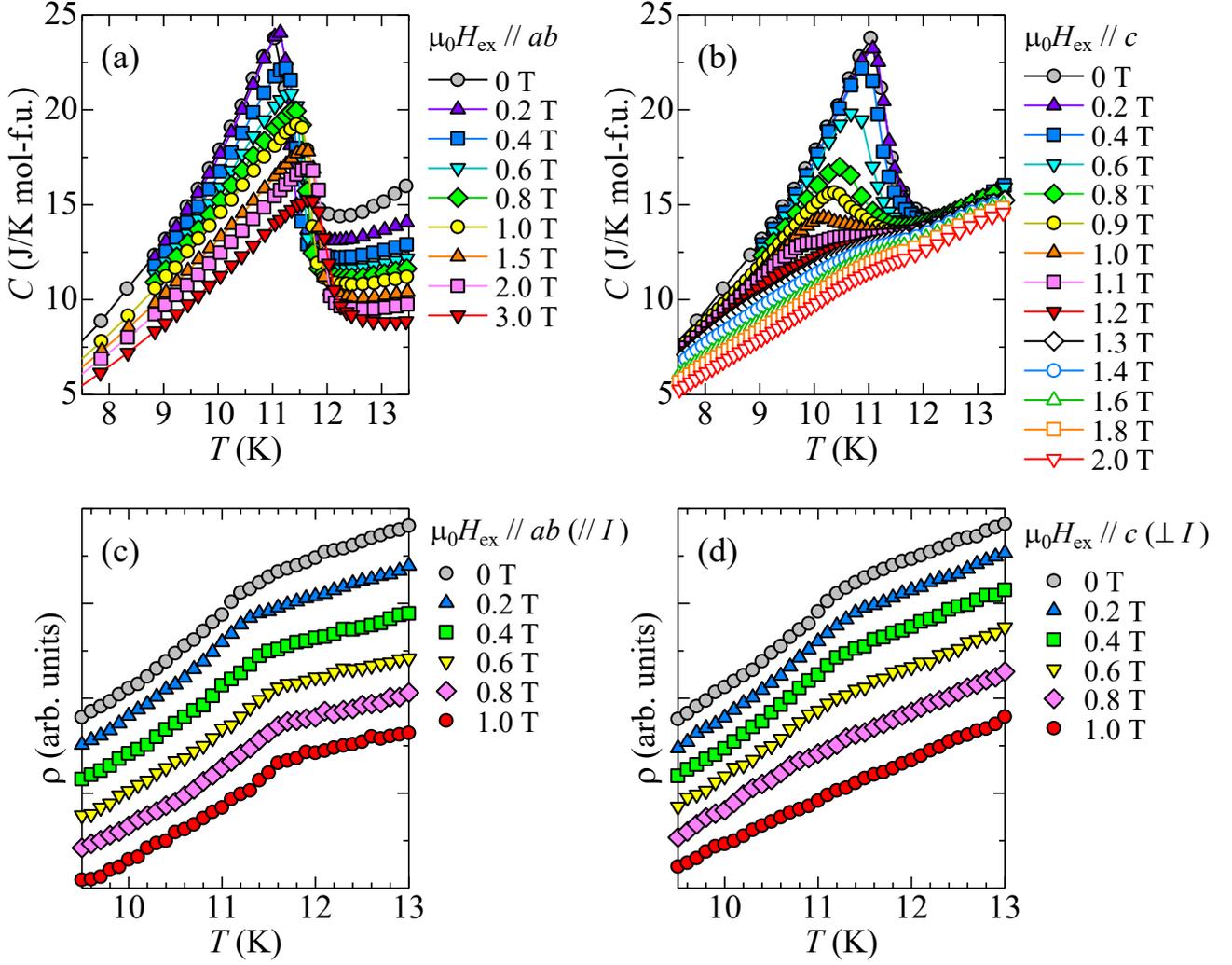

Fig. S5. (a)–(b) Specific heat peak around 11 K under various external magnetic fields, measured for (a) $\mu_0 H_{ex}$ // $ab$ and (b) $\mu_0 H_{ex}$ // $c$. (c)–(d) Temperature dependence of resistivity around 11 K under various external magnetic fields up to 1 T measured for (c) $\mu_0 H_{ex}$ // $ab$ and (d) $\mu_0 H_{ex}$ // $c$.

## V. MAGNETIC FIELD DEPENDENCE OF SPECIFIC HEAT AND ELECTRICAL RESISTIVITY NEAR THE PM–FM PHASE TRANSITION

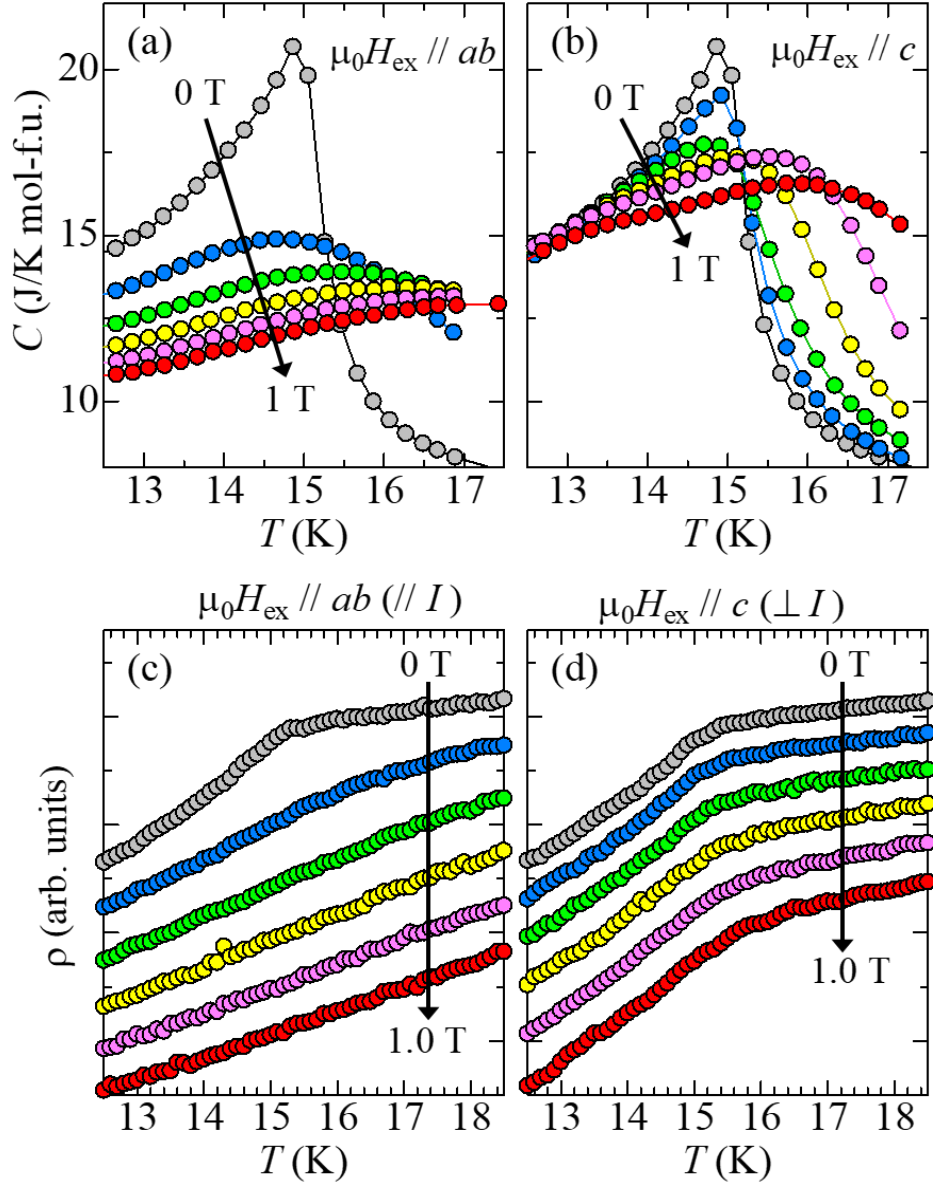

Fig. S6. (a)–(b) Specific heat peak around 15 K under various external magnetic fields up to 1 T, measured for (a) $\mu_0 H_{ex}$ // $ab$ and (b) $\mu_0 H_{ex}$ // $c$. (c)–(d) Temperature dependence of resistivity around 15 K under various external magnetic fields up to 1 T measured for (a) $\mu_0 H_{ex}$ // $ab$ and (b) $\mu_0 H_{ex}$ // $c$. The $\mu_0 H_{ex}$ increments are 0.2 T in (a)–(d).

## VI.  DETERMINATION OF THE MAGNETIC PHASE DIAGRAM

To obtain the magnetic phase diagram for $HoB_2$, the transition temperatures, $T_{low}$ and $T_C$, were determined from the temperature dependence of $C$ and $\rho$ under various external magnetic fields. Here, we defined the transition temperature as the temperature at which $C/T$ and $\rho/T$ show peaks. At zero magnetic field, we obtained $T_C$ and $T_{low}$ to be ~15 K and ~11 K, respectively.

Although $T_C$ cannot be defined at any magnetic field, a crossover line separating FM and PM regimes can be evaluated from the shift of specific heat peak near $T_C$, which is broadened by applying a magnetic field. Hence, we determined the crossover line by tracing the peak temperature of broad specific heat peak at each external magnetic field [see Fig. 3(a), Fig. 3(b), Fig. S6(a), and Fig. S6(b).].

$T_{low}$ was estimated from $C/T$ up to 3 T and $\rho/T$ up to 5 T for $\mu_0 H_{ex}$ // $ab$, as indicated by the black vertical bars in Fig. S7. In case of $\mu_0 H_{ex}$ // $c$, $T_{low}$ was estimated from $C/T$ up to 1.4 T [Figs. S8(a) and S8(b)] and $\rho/T$ up to 0.9 T [Fig. S8(c)]. Moreover, we determined the critical magnetic field from the magnetic field dependence of resistivity ($\rho$-$H$ curve) between 7.5 K and 10.5 K in the field range of 0.9–1.6 T. As shown in Fig. S8(d), $\rho$-$H$ curve for $\mu_0 H_{ex}$ // $c$ exhibits a plateau-like behavior, followed by a sudden drop with increasing a magnetic field. Notably, an inflection point of $\rho$-$H$ curve at 10.5 K, i.e., 0.90 T, coincides with the magnetic field where $\rho/T$ peaks at 10.5 K. Accordingly, we defined the inflection point at each temperature as the critical magnetic field for the low-temperature phase transition, as indicated by the black vertical bars in Fig. 8(d).

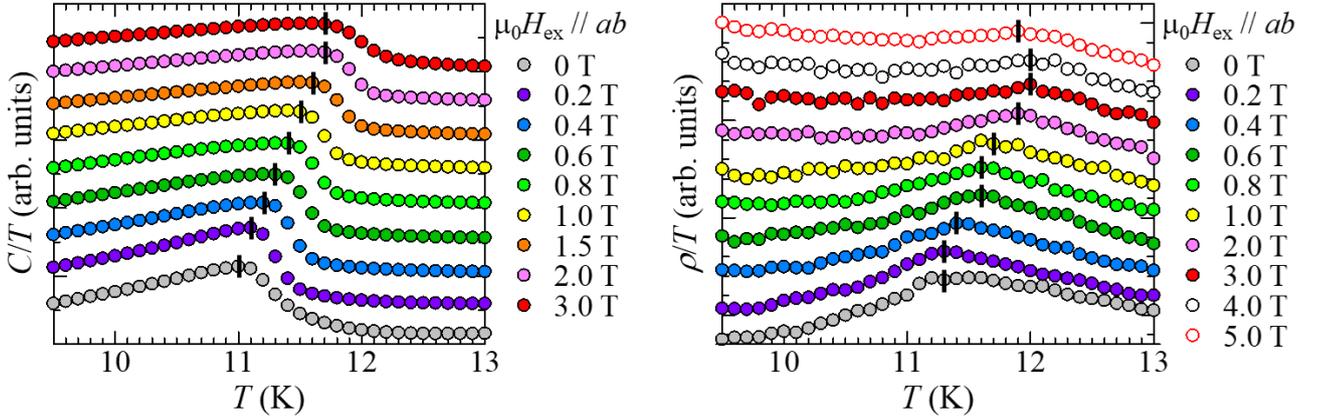

Fig. S7. Temperature dependence of $C/T$ (the left panel) and $\rho/T$ (the right panel) near 11 K under various external magnetic fields for $\mu_0 H_{ex}$ // $ab$. The black vertical bars in each figure represent the $T_{low}$ defined as the peak temperature of $C/T$ and $\rho/T$, respectively.

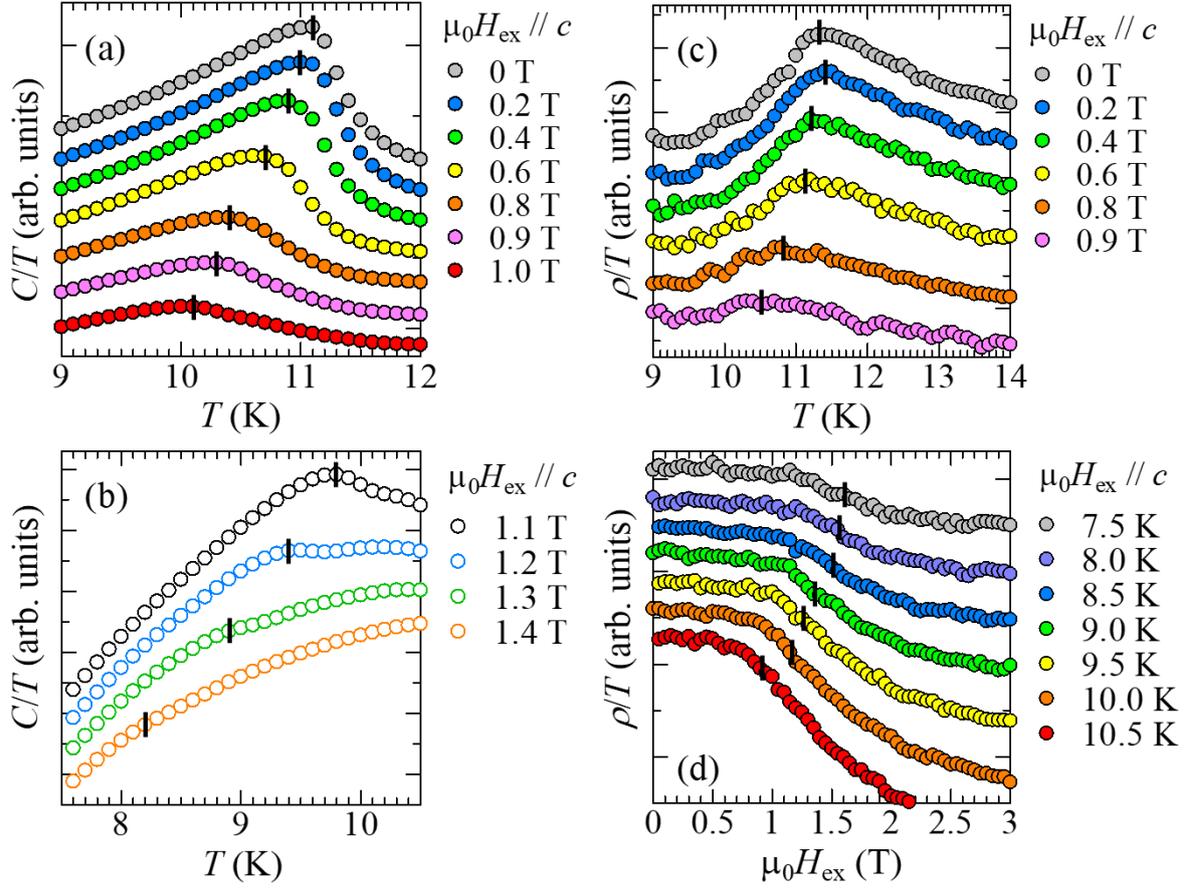

Fig. S8. (a)–(b) Temperature dependence of $C/T$ near 11 K under various external magnetic fields from 0 to 1.0 T (a) and from 1.1 to 1.4 T (b) for $\mu_0 H_{ex} // c$. (c) $\rho/T$ as a function of temperature under the magnetic fields up to 0.9 T. The black vertical bars in (a)–(c) represent $T_{low}$ defined as the peak temperature of $C/T$ and $\rho/T$, respectively. (d) Magnetic field dependence of $\rho$ ($\rho$-$H$ curve) at various temperatures between 7.5 and 10.5 K. The black vertical bars in (d) represent the critical magnetic fields, defined as the inflection point of $\rho$-$H$ curve at each temperature.

## VII. THERMAL HYSTERESIS OF ISO-FIELD MAGNETIZATION CURVES

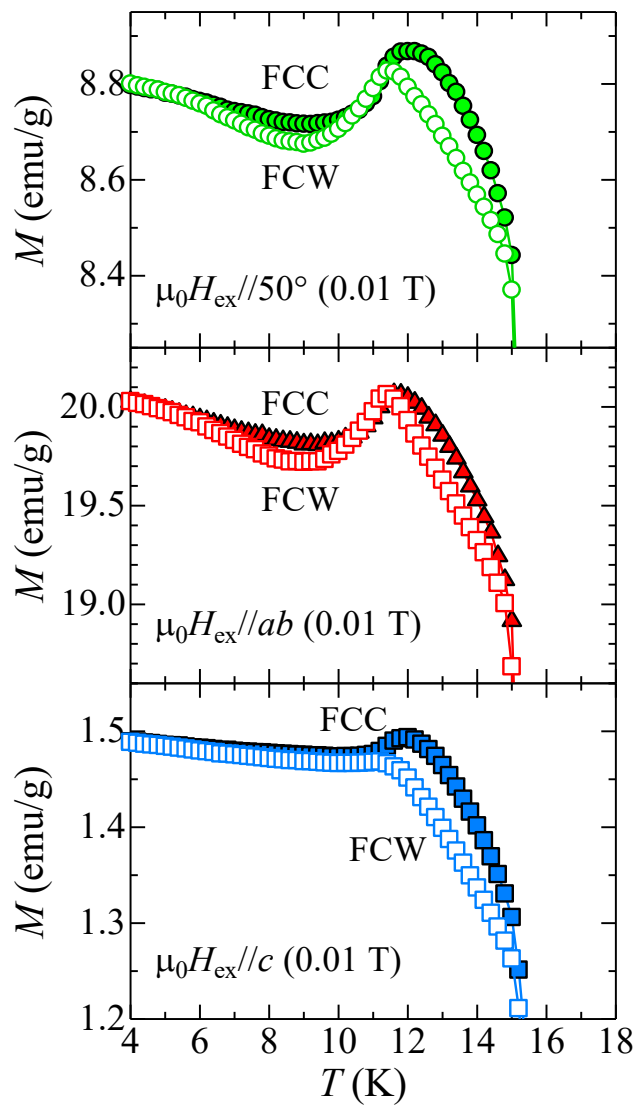

Fig. S9. Iso-field magnetization curves of the HoB$_2$ single crystal measured at 0.01 T under field-cooled-cooling (FCC) and field-cooled-warming (FCW) processes.